# Numerical studies of an eccentric tube-in-tube helically coiled heat exchanger for IHEP-ADS helium purification system

## Zhang Jianqin[1, 2;1] Li Shaopeng[1]

[1]Institute of High Energy Physics, Chinese Academy of Science, Beijing, 10049, China

[2] University of Chinese Academy of Sciences, Beijing 100049, China

**Abstract**: The tube-in-tube helically coiled (TTHC) heat exchanger is preferred in the purifier of IHEP-ADS helium purification system. The position of an internal tube is usually eccentric in a TTHC heat exchanger in practice, while most TTHC heat exchangers in the literature studied are concentric. In this paper, TTHC heat exchangers with different eccentricity ratios are numerically studied for turbulent flow and heat transfer characteristics under different flow rates. The fluid considered is helium at the pressure of 20Mpa, with temperature dependent thermo-physical properties for the inner tube and the annulus. The inner Nusselt number between the concentric and eccentric TTHC heat exchangers are compared, so is the annulus Nusselt number. The results show that with the eccentricity increasing, the annulus Nusselt number increases substantially. According to the numerical data, new empirical correlations of Nusselt number as a function of Reynolds number and eccentricity for the inner tube and the annulus are presented in this paper. The TTHC heat exchanger of the purifier is designed under the given conditions.

## 1. Introduction

Helium Purification System is an important sub-system which is under construction to meet the high helium purity standard of the cryogenic system in the Accelerator Driven Sub-critical System of the Institute of High Energy Physics (IHEP-ADS). The purifier includes dewar, high pressure heat exchanger, condenser, liquid air separator and adsorption cylinders. The dewar is fitted with a quantity of liquid nitrogen, in which condenser, liquid air separator and adsorbent cylinders are submerged. The high pressure heat exchanger is used to transfer the heat of helium between the temperature of 300K and 78K. Because of the high pressure of 20Mpa and the compact structure, tube-in-tube helically coiled (TTHC) heat exchanger is preferred.

Helically coiled heat exchangers are superior to straight tubes, for which have compact structures and higher heat transfer coefficients. It is known that centrifugal forces caused by the pipe curvature will result in a secondary flow, which enhances heat transfer rates. A number of studies focused on the fluid flow and heat transfer characteristics in helical pipes [1, 2, 3]. Mori and Nakayama [4] theoretically and experimentally studied the flow and heat transfer in a curved tube both in laminar and turbulent regions, and provided empirical correlations. Furthermore，Rennie and Raghavan[5] conducted experimental study of a TTHC heat exchanger with different curvature ratios. Numerical methods were used to study the helical tubes with Computational Fluid Dynamics (CFD) simulation [6]. Rennie and Raghavan [7] and Later Kumar [8] numerically studied the laminar and the turbulent flow and heat transfer of a TTHC heat exchanger. The literature above-mentioned mainly focused on the concentric TTHC heat exchangers, while in many practical situations, they are eccentric, coiled without sharing the same vertical or horizontal center line. Louw and Meyer [9] experimentally compared the annular contact TTHC heat exchanger with a concentric one. It's performed experimental study of natural convection and turbulent forced convection flow in vertical eccentric annulus [10-13].

In this work, it is proposed to study the fluid flow and heat transfer characteristics in concentric and vertical eccentric TTHC heat exchangers with CFD methods, and the eccentricity (e/ ($R_0$-$R_i$)) of the inner tube ranges from 0 to 1. Subsequently comparisons of the flow and heat transfer characteristics of the inner tube and the annulus will be conducted. New empirical correlations of Nusselt number versus Reynolds number and eccentricity for the inner tube and the annulus in eccentric TTHC heat exchangers are presented. The TTHC heat exchanger of the purifier is designed under the given conditions

## 2. Materials and methods

A sectional view of a vertical eccentric TTHC heat exchanger is show in Fig.1a. The eccentricity (e/ ($R_0$-$R_i$)) of the inner tube ranges from the concentric case (E=0) to the annular contact case (E=1), that is 0, 0.4, 0.8, 0.96 and 1 separately. Considering the manufacturing condition, the case of E=1 is welded with copper, which is shown in Fig.1b. Except for the different eccentricity ratios, the geometries of the eccentric TTHC heat exchangers are same. The outside diameter of the inner tube is 8mm with a wall thickness of 1mm, while the outside diameter of the outer tube is 18mm with a wall thickness of 2.5mm.

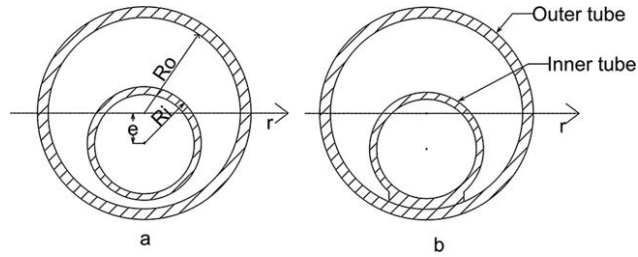

Fig.1. Sectional views of a TTHC heat exchanger (a) eccentric and (b) E=1.

The transport and thermal properties of helium were dependent on the temperature. FLUENT 14.0 in ANSYS was used in the simulation. The geometrical and flow parameters for the inner and outer tubes are given in table 1.The working fluid in the inner tube and the annulus was helium. These two fluids were counter flow and heat was transferred through the copper wall. The total number of simulations performed for each type of TTHC heat exchanger was 21 (11 inner tube flow rates and 10 annulus flow rates).

**Table 1**

Geometrical and flow parameters for the inner and outer tubes

|  | Inner tube | Outer tube |
|---|---|---|
| **Outer diameter(m)** | 0.008 | 0.018 |
| **Inner diameter(m)** | 0.006 | 0.013 |
| **Coil diameter(m)** | 0.3 | 0.3 |
| **Pitch(m)** | 0.021 | 0.021 |
| **Number of turns** | 2 | 2 |
| **Inlet temperature(K)** | 78 | 300 |
| **Pressure(bars)** | 198 | 200 |
| **Flow rate 1 (kg/s)** | 0.001-0.007 | 0.002 |
| **Flow rate 2 (kg/s)** | 0.002 | 0.0025-0.012 |

## 3. Results and discussion

### 3.1 Temperature profiles

The temperature profiles at the exit of the inner tube and the annulus for different eccentricity ratios of TTHC heat exchangers are given in Fig.2. The mass flow rate in the inner tube is 2 g/s, and the mass flow rate in the annulus is 10 g/s. In the Fig.2a, the outlet temperature profiles of the inner tube for different eccentricities are similar and the average temperature decreases with E increasing. The right side is the inner side of the coil, and has a higher temperature compared to the outer side. The average outlet temperature for the inner tube of the concentric heat exchanger is higher than the eccentric ones. In the Fig.2b, the right side is the inner side of the coil. With the center offset increasing, the high temperature region moves from the outer side to the inner side. Moreover, the secondary flow enhanced the heat transfer in the annulus region with E increasing.

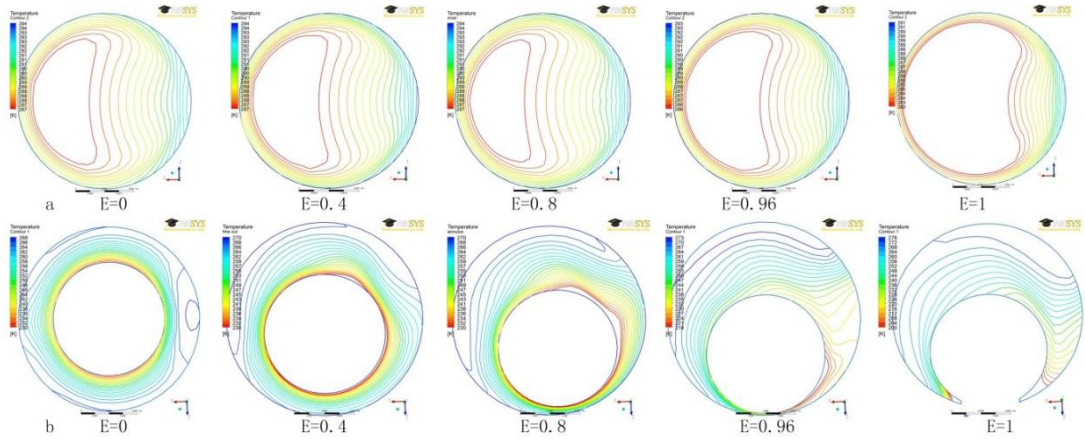

Fig.2. Temperature profiles in the inner tube /the annulus for different eccentricities: (a) the inner tube and (b) the annulus.

### 3.2 Inner tube and annulus heat transfer coefficient

Heat transfer coefficients for the inner side, $h_i$ and the annulus tube side, $h_o$, were calculated with traditional Wilson plots [3, 5]. Wilson plots allow the heat transfer coefficients to be calculated based on the overall heat transfer coefficient, without the requirement of wall temperature. The outer Reynolds number of the annulus was calculated by equivalent diameter. New empirical models were developed with Sieder-Tate type [14, 15] equation, which was used to fit the inner and annulus Nusselt number with a coiled effect term $\delta^{0.1}$ [4] and was the function of eccentricity. The differences between the empirical data and the numerical data are less than 2%. The equations are:

$$Nu_i = C_i \operatorname{Re}_i^{t_i} \operatorname{Pr}_i^{\frac{1}{3}} (\frac{\mu_i}{\mu_w})^{0.14} \delta_i^{0.1} \tag{1}$$

$$Nu_o = C_o \operatorname{Re}_o^{t_o} \operatorname{Pr}_o^{\frac{1}{3}} (\frac{\mu_o}{\mu_w})^{0.14} \delta_o^{0.1} \tag{2}$$

where the values of the coefficients $C_i$, $C_o$ and $t_i$, $t_o$ are given in Table 2, for 14,000<$Re_i$<124,000, 8,500<$Re_o$<37,000.

**Table 2**

Values of the coefficients for the inner and the annulus Nusselt number equation in the concentric and eccentric TTHC heat exchangers

| E | 0 | 0.4 | 0.8 | 0.96 | 1 |
|---|---|---|---|---|---|
| $C_i$ | 0.1907 | 0.2414 | 0.4040 | 1.4354 | 3.0790 |
| $t_i$ | 0.6428 | 0.6096 | 0.5591 | 0.4408 | 0.3771 |
| $C_o$ | 0.4688 | 0.2435 | 0.0926 | 0.0327 | 0.0186 |

| $t_o$ | 0.4824 | 0.5589 | 0.6704 | 0.7797 | 0.8411 |

The empirical constants ($C_i$, $t_i$, $C_o$, $t_o$) are related to $E$, and the relations are as follows:

$$C_i = 2.807E^{21.219} + 0.268 \qquad (3)$$

$$t_i = -0.399E^2 + 0.173E + 0.634 \qquad (4)$$

$$C_o = -0.45E + 0.441 \qquad (5)$$

$$t_o = 0.325E^{2.436} + 0.498 \qquad (6)$$

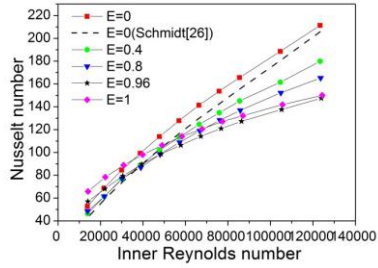

Fig.3. Nusselt number vs. inner Reynolds number, the average annulus Reynolds number is 7000.

Fig.3 illustrates that the inner Nusselt number decreases by approximately 11.8%, 14.4%, 17.7% and 15.6% for E=0.4, E=0.8, E=0.96 and E=1 compared with E=0. The reason is probably that the eccentricity causes the temperature non-uniformity of the wall, and heat transfer turns worse in the lowering wall temperature region compared with the concentric one. Fig.3 also shows that the experimental data of Schmidt [16] has a good agreement with the concentric TTHC heat exchanger.

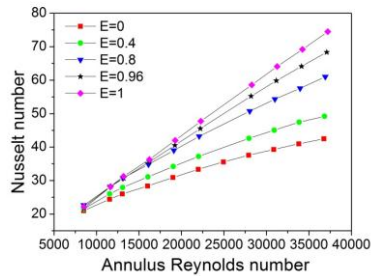

Fig.4. Nusselt number vs. annulus Reynolds number, the average inner Reynolds number is 28000.

Fig.4 shows the Nusselt number increases with the annulus Reynolds number increasing. It can be seen from the figure that the annulus Nusselt number increases with E increasing, especially at the large Reynolds number. The Nusselt numbers in the cases of E=0.4, E=0.8, E=0.96, and E=1 are approximately 10.8%, 27.8%, 34.7% and 41.2% separately higher than those in the case of E=0. With the eccentricity ratio increasing, the flow non-uniformity is enhanced with secondary flow and the heat transfer is improved the in the annulus.

The relationship between the overall heat transfer coefficient and the inner /annulus heat transfer coefficient was as follows:

$$\frac{1}{U_o} = \frac{A_o}{A_i h_i} + R_c + \frac{1}{h_o} \qquad (7)$$

Compared with the inner and annulus heat transfer coefficients, the conduction resistance term $R_c$ is negligible. It is obvious in the Fig. 3 and Fig. 4 that the Nusselt number in the annulus is less than the inner tube at the same mass flow rate. So an increase of eccentricity is good for the increase the overall heat transfer coefficient.

### 3.3 Friction factors

The inner and annulus friction factors were as follows:

$$f = \frac{2\Delta P \times D_e}{\rho u_0^2 \times L} \tag{8}$$

where $\Delta P$ is the pressure drop of the inner tube and the annulus, $L$ is the length of the heat exchanger, $D_e$ is the equivalent diameter of the inner tube and the annulus and $u_0$, r are the average velocity and the density of the inner tube and the annulus, respectively.

Fig.5 shows the numerical data in the inner tube and the annulus for the concentric and eccentric TTHC heat exchangers and the experimental data of Schmidt [16]. Fig.5a presents that the inner friction factors of the concentric and eccentric TTHC heat exchangers are almost same at the same Reynolds number, and have a good agreement with Schmidt's experimental data. Fig.5b presents that the annulus friction factor has a decrease of 13% and 25% for E=0.96 and E=1 separately compared with the value of E<0.96.

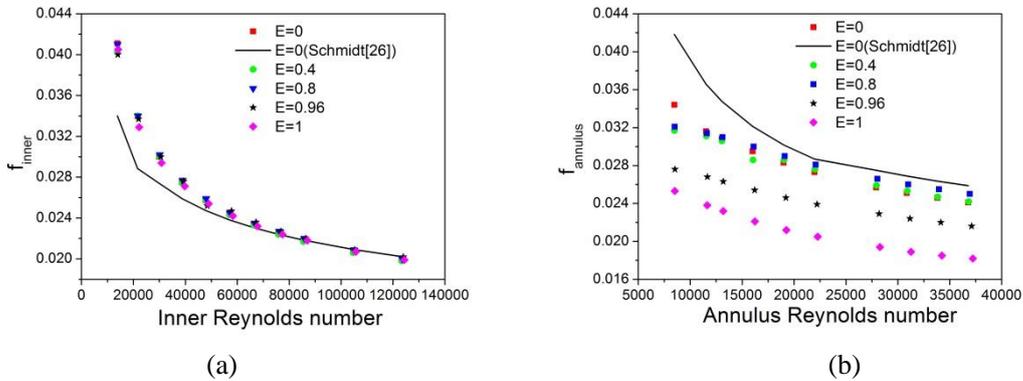

(a)                                             (b)

Fig.5. Friction factor versus (a) inner Reynolds number and (b) the annulus Reynolds number.

### 3.4. Design of the heat exchanger

According to the analysis in 3.2, the case of E=1 TTHC heat exchanger is preferred to design the heat exchanger in the purifier, especially in the high Reynolds number region. The design parameter of the heat exchanger is shown in Table 3. The diameter of the inner tube is 8mm with a thickness of 1mm, and the diameter of the outer tube is 16mm with a thickness of 2.5mm. The coil diameter is 300mm. The inner and annulus heat transfer coefficients are calculated with Equations (1) and (2). The length of the tube is 26 m. The pressure drop of the inner tube and the annulus are 0.3bar and 0.4bar separately.

Table 3. Parameters of the helically coiled tube-in-tube heat exchanger

|  | Input temperature K | Input Pressure MPa | Flow rate g/s | Output temperature K |
|---|---|---|---|---|
| Input gas | 300 | 20 | 5.2 | 93.5 |
| Output gas | 78 | 19.8 | 4.9 | 295 |

## 4. Conclusions

Numerical studies of different eccentricity ratios of TTHC heat exchangers were performed with helium in the inner tube and the annulus. With the eccentricity increasing, the inner Nusselt number decreases slowly, while the annulus Nusselt number increases rapidly. In the case of E=1, the inner Nusselt number decreases by approximately 15.6% and the annulus Nusselt number increases by approximately 41.2% compared with the case of E=0. Eccentricity is good for the heat transfer of the annulus in an eccentric TTHC heat exchanger, especially in the large Reynolds number region. New empirical correlations of the Nusselt number versus the Reynolds number and eccentricity in the inner tube and the annulus were presented. The friction factors of the inner tube and annulus were studied for the concentric and eccentric TTHC heat exchangers. The heat exchanger of the purification system was designed with the case of E=1 TTHC heat exchanger. In order to examine the reliability of CFD results, an experiment for an eccentric TTHC heat exchanger should be conducted in the future.

## References

[1]   R.A. Seban, E.F. McLaughlin, Heat transfer in tube coils with laminar and turbulent flow, International Journal of Heat and Mass Transfer, 1963, 6: 387–395.

[2]   G.F.C. Rogers, Y.R. Mayhew, Heat transfer and pressure loss in helically coiled tubes with turbulent flow, International Journal of Heat and Mass Transfer, 1964, 7: 1207–1216.

[3]   Patankar SV, Pratap VS, Spalding DB. Prediction of laminar flow and heat transfer in helically coiled pipes. J Fluid Mech, 1974, 62: 53–551.

[4]   Y. Mori, W. Nakayama, Study on forced convective heat transfer in curved tubes, (3rd report, theoretical analysis under the conditions of uniform wall temperature and practical formulae), International Journal of Heat and Mass Transfer, 1967, 10: 681–695.

[5]   T.J. Rennie, G.S.V. Raghavan, Experimental studies of a double pipe helical heat exchanger, Experimental Thermal and Fluid Science, 2005. 919–924.

[6]   M. M. Asiam Bhutta, N. Hayat, M. H. Bashir et.al, CFD applications in various heat exchangers design: A review, Applied Thermal Engineering, 2012, 32: 1–12.

[7]   T.J. Rennie, G.S.V. Raghavan, Numerical studies of a double pipe helical heat exchanger, Applied Thermal Engineering, 2006, 26: 1266–1273.

[8]   V. Kumar et al, Numerical studies of a tube-in-tube helically coiled heat exchanger, Chemical Engineering and Processing, 2008, 47: 2287–2295.

[9]   Willem I. Louw ,Josua P. Meyer, Heat Transfer during Annular Tube Contact in a Helically Coiled Tube-in-Tube Heat Exchanger, Heat Transfer Engineering, 2005, 26(6): 16–21.

[10]  Hosseini R, Heyrani-Nobari M R, Hatam M. An experimental study of heat transfer in an open-ended vertical eccentric annulus with insulated and constant heat flux boundaries, Applied thermal engineering, 2005, 25(8): 1247–1257.

[11]  Hosseini R, Rezania A, Alipour M, et al. Natural convection heat transfer from a long heated vertical cylinder to an adjacent air gap of concentric and eccentric conditions, Heat and Mass Transfer, 2012, 48(1): 55-60.

[12]  Hosseini R, Ramezani M, Mazaheri M R. Experimental study of turbulent forced convection in vertical eccentric annulus, Energy Conversion and Management, 2009, 50(9): 2266-2274.

[13]  Shiniyan B, Hosseini R, Naderan H. The effect of geometric parameters on mixed convection in an inclined eccentric annulus, International Journal of Thermal Sciences, 2013, 68: 136-147.

[14]  J.W. Rose, Heat-transfer coefficients, Wilson plots and accuracy of thermal measurements, Experimental Thermal and Fluid Science, 2004, 28: 77–86.

[15]  Incropera, F.P., and DeWit, D. P., Introduction to Heat Transfer, 3$^{rd}$ ed., New York: John Wiley & Sons, 1996, p. 544.

[16]  Schmidt, E.F., Warmeubergang and Druckverlust in rohrschlangen, Chemie-Ing-Techn, 1967, 39:781-789.